# Characterization of active matter in dense suspensions with heterodyne laser Doppler velocimetry.

Johannes Sachs[1], S. Nikhilesh Kottapalli[1], Peer Fischer[1,2], Denis Botin[3], Thomas Palberg[3]

[1] Max Planck Institute for Intelligent Systems, Heisenbergstr. 3, 70569 Stuttgart

[2] Institute of Physical Chemistry, University of Stuttgart, Pfaffenwaldring 55, 70569 Stuttgart

[3] Institute of Physics, Johannes Gutenberg University, Staudinger Weg 7, 55128 Mainz

**Abstract**

We present a novel approach for characterizing the properties and performance of active matter in dilute suspension as well as in crowded environments. We use Super-Heterodyne Laser-Doppler-Velocimetry (SH-LDV) to study large ensembles of catalytically active Janus particles moving under UV-illumination. SH-LDV facilitates a model-free determination of the swimming speed and direction, with excellent ensemble averaging. In addition we obtain information on the distribution of the catalytic activity. Moreover, SH-LDV operates away from walls and permits a facile correction for multiple scattering contributions. It thus allows for studies of concentrated suspensions of swimmers or of systems where swimmers propel actively in an environment crowded by passive particles. We demonstrate the versatility and the scope of the method with a few selected examples. We anticipate that SH-LDV complements established methods and paves the way for systematic measurements at previously inaccessible boundary conditions.

**Introduction**

Active Matter is a rapidly evolving emergent field of Soft Matter Physics triggering intense experimental and theoretical activity [1-3]. Catalytic Janus particles present a model system within the large class of self-propelled particles [4-17]. Properties and performance of individual particles and small ensembles are readily accessible by (confocal or holographic) microscopy utilizing particle tracking and image analysis for structural and dynamical characterization [18-20]. Particle image velocimetry [21] or Dynamic Differential Microscopy (DDM) are suitable for studies of larger ensembles of individually propelling swimmers [17,22,23]. Both approaches, however, reveal severe technical drawbacks (e.g. low statistics or the necessity for a fitting model), when more interesting situations are addressed. Swimming in crowded environments, collective swimming and density dependent phase separation have already received a lot of theoretical interest [1,2,24-30], which, however is not yet fully complemented by experimental work. The main experimental challenges are reliable ensemble averaging and increased turbidity of concentrated or crowded samples. In particular, multiple scattering severely complicates acquisition of reliable information from the collected data in many optical approaches at large densities. Therefore, Multi-speckle X-ray photon correlation spectroscopy (XPCS) has been applied which is able to return information on active motion fluctuations and passive diffusion simultaneously [31]. This however, is instrumentally somewhat more demanding than cross correlation techniques in conventional homodyne or heterodyne laser light scattering approaches [32,33]. Moreover, correlation schemes based on homodyne scattering return information on fluctuations of the number of particles in the probed volume [34], or on the velocity differences within the probed volume as projected on the scattering vector $q$ [35], which both relate to the velocity fluctuations about its mean. In particular the latter information may be very difficult to extract in situations, where the velocity differences within the probed volume do not remain small [36]. Correlation techniques can therefore separate the advective and diffusive components of the dynamics along vertical and horizontal directions, but not return magnitudes and directions of average velocities. In the present paper, we present a simple alternative for characterizing the swimming performance of catalytic Janus nanoparticles in terms of their average velocities and directions. We based our approach on Laser Doppler Velocimetry (LDV), which is a heterodyne light scattering technique analysing data in frequency domain instead of correlation time domain. Use of this approach has been reported already very early, for studies of particle diffusion [37] but it was not much followed after the

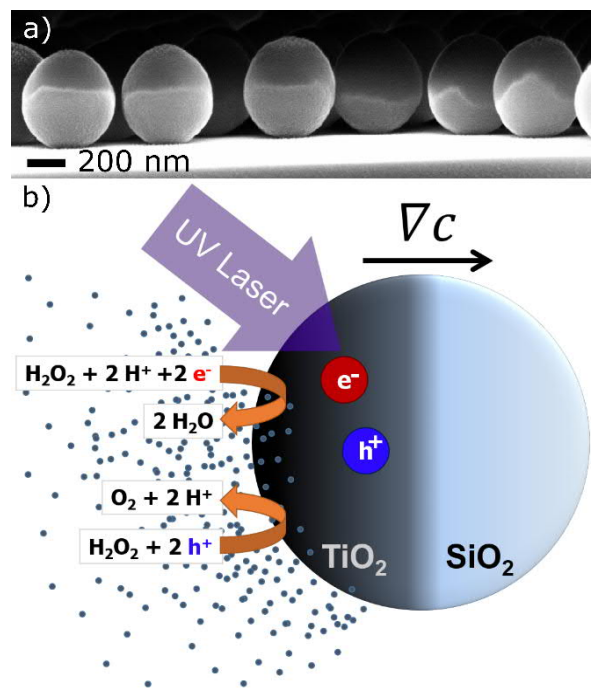

**Fig. 1** a) SEM image of the as prepared $SiO_2/TiO_2$ Janus particles JP550 (scale bar 200nm). b) Schematics of the photo-catalytically active Janus particle. Upon UV irradiation electron-hole pairs are generated in the anatase-$TiO_2$ that will drive the decomposition of $H_2O_2$. The decomposition occurs only on one half of the particle, thus giving rise to a concentration gradient across the swimmers surface that will move the particle self-phoretically through the surrounding fluid [14].

rapid development of time domain Dynamic Light Scattering (DLS). However, it quickly became a versatile standard approach to probe flowing, sheared or even turbulent systems [38-40].

We here employ the recently introduced Super-Heterodyne Laser Doppler-Velocimetry (SH-LDV) allowing for the correction of multiple scattering by an empirical post-recording correction scheme [41]. There, both velocimetry and the determination of average diffusion coefficients, D, was possible with a statistical uncertainty of $\sigma \leq \Delta D/D \approx 0.02$ at transmissions of 40% and larger and we studied systems undergoing phase transitions and suspensions under symmetric and asymmetric solvent flows [42]. For the competing cross correlation techniques, similar values have been reported, with modulated 3D cross correlation showing somewhat better marks due to the modulation induced increase of the intercept [32]. In the instrumental setup used here, we combine super-heterodyning (rendering the actual signal free of low-frequency noise and artefacts) with small angle scattering (allowing model-free access to the velocity distribution irrespective of the sample structure) and a large detection volume (providing excellent ensemble averaging). Moreover, we use a facile frequency-space correction scheme to isolate and subtract the multiple-scattering background and obtain high quality Doppler spectra for evaluation. This for the first time enables us to characterize the performance of catalytic self-propelled micro-swimmers in a turbid background bulk suspension of passive particles.

**Experimental**

For our demonstrations, we chose $SiO_2/TiO_2$ Janus particles based on a $SiO_2$ bead with hydrodynamic radius $a_h$ = 259±7 nm (SiO2-F-SC68, microParticles GmbH, Berlin, Germany) onto which $TiO_2$ has been deposited. The Janus particles are fabricated similar to the ones described in [14]. Briefly, a closed-packed monolayer of $SiO_2$ beads is formed using the Langmuir-Blodgett technique and then transferred onto a Si-wafer, followed by a physical vapour deposition of $TiO_2$ under an oblique angle (85°) and continuous (fast) rotation of the substrate. Afterwards the sample is annealed (450°C for 2h) in order to form anatase-$TiO_2$ and subsequently dispersed into solution via a sonication bath. The dry radius after coating is $a_{SEM}$ = (275±8) nm. The lab code for these particles is therefore JP550 where JP denotes the Janus Particle and the number denotes the Scanning Electron Microscope (SEM) diameter. Fig. 1 a) shows an SEM image of the $TiO_2$-capped Janus particles on the wafer. Our fabrication scheme allows to manufacture a full wafer yielding a high number of particles ($N\sim10^{10}$) at once, giving us the freedom to create highly concentrated particle suspensions that are necessary to perform bulk light scattering experiments with larger sample volumes.

Under excitation with UV light (λ=375nm) electron-hole pairs can be generated in the anatase-$TiO_2$ cap which activate the decomposition of $H_2O_2$, causing a chemical gradient across the Janus particle surface. This gradient drives the self-phoretic motion as schematically depicted in Fig. 1b). In the present experiments, the peroxide concentrations ranged from 0.5% to 3%. By switching the illumination on and off, we can change the Janus particles' motion from purely Brownian passive to self-propelling active swimming. This type of catalytically active Janus particles are able to follow the direction of the incident UV light, and can even propel against gravity when illuminated from below [14]. Low concentration samples of Janus particles in general [6,10] and of light-switchable types in particular have been extensively investigated elsewhere [7,9,12,14]. However, the vast majority of experiments are measured in close proximity of a surface, which could dramatically influence the propulsion [5] and interaction [13] characteristics. Moreover, most studies rely on particle tracking being subject to

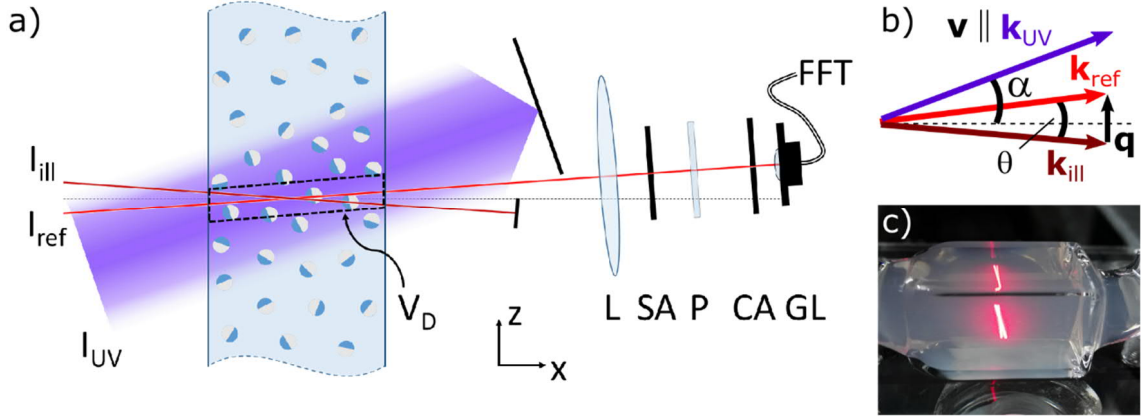

**Fig. 2** a) Experimental setup and scattering geometry in top-view. Illuminating ($I_{ill}$, $\omega_{ill} = \omega_0$) and reference ($I_{ref}$, $\omega_{ref} = \omega_0 + \omega_{Bragg}$) beams, cross under a scattering angle $\theta_S = 7.6°$ inside the suspension filled sample cell with thickness d = 10 mm (refraction at cell walls not drawn here). The reference beam is collinear with the observation direction. Scattered light is focused by lens (L) at f = 50 mm onto a small circular aperture (CA), collected by a gradient-index lens (GL) and fed into an optical fibre leading to the detector (not shown). Precise location of CA selects the detected vector of scattered light and restricts it to $\mathbf{k}_s = \mathbf{k}_{ref}$. The distance CA-GL defines the diameter of the cylindrical detection volume ($V_D$, boxed area inside the cell) adjusted to contain the complete path of the illuminating beam inside the cell. Note that it covers the central region of the widened UV beam activating the catalytic swimmers. A horizontal slit aperture (SA) rejects any light travelling outside the *x-z* plane and a polarizer (P) assures V/V scattering geometry. b) Wave vectors of the illumination beam, $\mathbf{k}_{ill}$, and of the reference beam, $\mathbf{k}_{ref} = \mathbf{k}_s$, yielding a scattering vector $\mathbf{q}$ parallel to the cell axis. The scattered light is Doppler shifted by $\omega_D = \mathbf{q} \cdot \mathbf{v}_z = (\mathbf{k}_s - \mathbf{k}_{ill}) \cdot \mathbf{v}_z$, which is positive, if the particles move co-linear with the wave vector of the UV illumination, $\mathbf{k}_{UV}$. c) A photograph of the sample cell filled with dense suspension of beads make the pronounced multiple-scattering halo of the reference as well as the illumination beam visible so they can be seen by the naked eye.

serious statistical issues[18,19]. It's therefore highly desirable to have alternative measurement techniques that are able to measure the Janus particle propulsion in bulk, avoiding perturbative wall effects and yielding a much better statistical confidence.

Crowded environments were examined by suspending the active Janus particles in a dispersion of passive polystyrene latex micro-particles of radius $a_h$ = 167 nm and size dispersity *PI*=0.08 (IDC Portland, USA, batch #1421, labcode PS310). Depending on the PS310 concentration this yielded samples with strongly reduced optical transmittance and substantial amounts of multiple scattering. Our systems are studied with an SH-LDV instrument (Fig. 2a) that has recently been described elsewhere with a detailed characterization of its performance[41]. The technique employs super-heterodyning [43] to separate the desired super-heterodyne part from homodyne and low frequency noise contributions to the measurement signal. In short, light of an illuminating beam and a reference beam propagate through a cuvette containing the sample. The reference beam and light scattered by the particles superimpose at the detector which monitors the resulting intensity. The frequency-shifted reference beam acts as a local oscillator generating a beat in intensity, which will be altered by the Doppler shift of the scattered light. A hardware based Fourier transform analyser (OnoSokki DS2000, Compumess, Germany) provides the power spectrum of the measured intensity time trace as a function of linear frequency $f = \omega / 2\pi$.

In general, the measured intensity contains parts stemming from the homodyne scattering (scattered light mixing itself) and from the heterodyne scattering (scattered light mixing with the light from the local oscillator). These contributions are separated in frequency space *via* a shift of the heterodyne part by $\omega_{Bragg}$. The shift between reference and illuminating beams is introduced utilizing a pair of stabilized acousto-optic modulators (AOM) delivering a nominal frequency difference of $\omega_{B,nominal} \pm \Delta\omega_{nominal} = 3000.0\pm0.5$ Hz. In practice, when setting the AOM, we found centre frequencies in the range: $\omega_{B,exp}(t=0) = \omega_{B,nominal} \pm 0.2$ Hz. We also checked for the temporal stability and found that over the duration of a typical measurement (30 min) $\omega_{B,exp}(t) = \omega_{B,exp}(t=0)\pm0.04$ Hz. The present SH-LDV instrument further features small angle scattering, which allows an efficient correction scheme to isolate the single-scattering signal from undesired multiple-scattering contributions [41] (for a recently realized prototype for wide angle scattering see[44])

The optical axis of the setup coincides with the propagation direction of the reference beam, i.e., $\mathbf{k}_{ref}$. The sample is further illuminated by a widened parallel UV laser ($I_{UV}$ = 19.5 mW) to UV-activate the catalytic particles. The UV illuminated region covers both reference and illumination beams, as shown in Fig. 2a). The propagation direction of the UV beam inside the cell encloses an angle $\alpha$ with the x-axis. Neglecting the small difference in refractive index for the two wavelengths, we typically chose $|\alpha|=18.8°$. For some experiments, we switched the direction of the UV illumination symmetrically with respect to the x- axis. In the present experiment, the detection optics ensure that only the light scattered with the same wave vector as of the reference beam $\mathbf{k}_s = \mathbf{k}_{ref}$ is collected. It is convenient to base the definition of the scattering vector **q** on the momentum transfer from the moving particles to the scattered photons:

$\mathbf{q} = \mathbf{k}_s - \mathbf{k}_{ill}$ pointing in positive z-direction [45]. Its magnitude $|\mathbf{q}| = q = \frac{4\pi n_S}{\lambda_0}\sin(\theta/2)$ depends on the scattering angle θ, the laser wave length $\lambda_0$ and the suspension refractive index $n_S$. In the present configuration we have λ = 633 nm, Θ = 7.6° and $n_S$ = 1.333, which results in q = 1.75 µm$^{-1}$. Note that the light scattered by particles moving in the UV illumination direction is Doppler shifted with a frequency:

$$\omega_D = \mathbf{q} \cdot \mathbf{v}(x,y,z) = qv_z = qv(x,y,z) \cdot \sin(\alpha) \qquad (1),$$

which depends on the component of the (spatially-dependent) velocity of the particles v(x,y,z) projected onto the scattering vector. Here α denotes the angle between the UV- and x-axis, and $v_z$ is the velocity's z-component, which is parallel to **q**. Motions out of the scattering plane (e.g. driven by thermal convection) are therefore not detectable in the present set-up. During a measurement, each individual frame covers a duration of $t_{frame}$ = 32 s, with overlap adjustable between 0 and 16 s. The signal gets averaged over successive frames for a total measurement time, T, typically a few thousand seconds. In a few runs, we found additional, transient intensity readings, which were significantly larger than average. This was traced back to bubble formation due to the catalytic processes at the particles. Corresponding runs were discarded.

A detailed scattering theory was already worked out elsewhere [46]. Briefly, we consider the case of scattered light with Gaussian statistics and particles drifting with a constant velocity **v**. In addition, we assume the particles to undergo Brownian motion with an effective diffusion coefficient, $D_{eff}$, which may depend on direct and hydrodynamic particle interactions, as well as on activity. The super-heterodyne power spectrum $C_{shet}(\mathbf{q},\omega)$, is the time Fourier transformation of the super-heterodyne mixed-field intensity autocorrelation function, $C_{shet}(\mathbf{q},\tau)$, that generally contains homodyne and heterodyne contributions:

$$C_{shet}(\mathbf{q},\omega) = \frac{1}{\pi}\int_{-\infty}^{\infty} d\tau \, \cos(\omega\tau) C_{shet}(\mathbf{q},\tau) \qquad (2)$$

Here ω is the circular frequency and τ the correlation time. In [46] we had derived its form in detail accounting for all components arising from singly scattered and reference light. Parasitic scattering at optical surfaces, multiple scattering and noise terms were not considered, as they are simply superimposing and can conveniently be removed from the raw data [41]. The corresponding single scattering power spectrum reads:

$$C_{shet}(\mathbf{q},\omega) = \left[I_{ref} + \langle I_s(\mathbf{q})\rangle\right]^2 \delta(\omega)$$
$$+ \frac{I_{ref}\langle I_s(\mathbf{q})\rangle}{\pi}\left[\frac{q^2 D_{eff}}{\left(\omega+[\omega_B+\omega_D]\right)^2+\left(q^2 D_{eff}\right)^2} + \frac{q^2 D_{eff}}{\left(\omega-[\omega_B+\omega_D]\right)^2+\left(q^2 D_{eff}\right)^2}\right] \quad (3)$$
$$+ \frac{\langle I_s(\mathbf{q})\rangle^2}{\pi}\frac{2q^2 D_{eff}}{\omega^2+\left(2q^2 D_{eff}\right)^2}$$

where $I_{ref}$ is the reference beam intensity, and $\langle I_s(q)\rangle$ is the time-averaged single scattering intensity for the chosen $q$. The subscript *shet* stands to specify the case of super-heterodyning, i.e. including the frequency shift between illuminating and reference light. The super-heterodyne power spectrum contains three terms: a trivial static background, the super-heterodyne Doppler signal and the homodyne signal (for examples of typical raw-spectra see e.g. [41,44,42]). Due to the super-heterodyning these terms are well separated in central frequencies. Both the static δ-peak and the homodyne signal are insensitive to the particle drift motion and are centred at $\omega = 0$. The two super-heterodyne Lorentzians are centred at $\pm(\omega_B+\omega_D)$, i.e. for a particle motion in the direction of UV illumination, resulting in a positive Doppler shift, both parts of the spectrum are shifted further outward from the origin. In the present integral mode, scattered light is collected from the complete cross section of the cell, which is illuminated by the central part of the widened UV beam that is activating the self-propulsion of the particles. The super-heterodyne signal thus averages over the projection of all velocities present in the detection volume. In fact, writing a normalized particle velocity distribution, $p(v) \sim dx/dv$ [47] in terms of the normalized distribution of Doppler frequencies $p(\omega_D)$, the spectrum can be written as convolution integral:

$$C_{shet}(\mathbf{q},\omega) = \int d\omega_D \; p(\omega_D) C^0_{shet}(\mathbf{q},\omega) \quad (4)$$

Both homodyne and background terms of Eq. 2, stay unaffected by the convolution with the particle velocity distribution. The two parts of the super-heterodyne Doppler spectrum, now, are diffusion-broadened distributions of Doppler frequencies about the averages, determined by the mean particle velocity. Note, that for any given UV illumination, the shape of $C_{shet}(\mathbf{q},\omega)$ is determined by a convolution of the intensity distribution across the detection volume with the distribution of catalytic activities. Thus, for dilute suspensions of large particles with weak Brownian motion and weak spatial variation of UV intensity, we should be able to obtain valuable information about the catalytic activity via the catalytic self-propulsion.

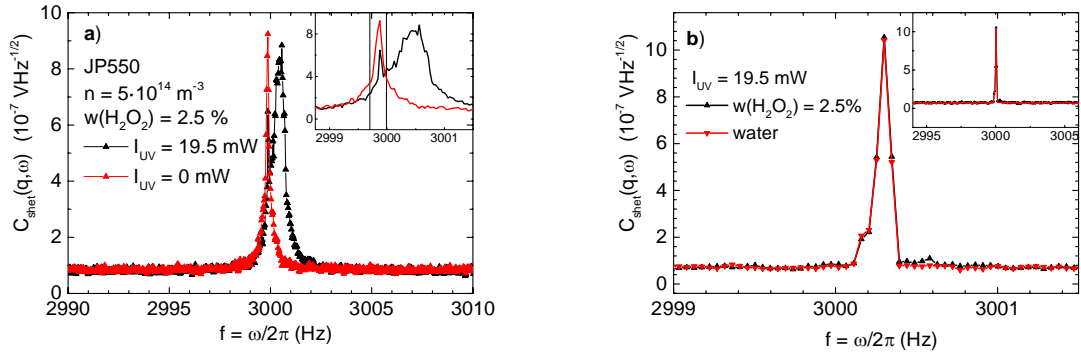

**Fig. 3** Typical spectra obtained by a SH-LDV experiment. a) Power spectra obtained by averaging over N = 54 individual frames (total measurement time T = 1500 s) for JP550 with w($H_2O_2$) = 2.5% without (red) and with (black) UV illumination. The spectrum for particles that are not catalytically active is a Lorentzian, centred at the Bragg frequency $\omega_B \approx 3$ kHz. Active motion towards the detector results in a positive Doppler shift. The signal change in shape reflects the diffusion broadened velocity distribution appearing under UV-illumination. Inset: magnified region close to $\omega_B$ Note that the Bragg frequency here becomes visible as sharp peak centred at $\omega_B$ = 2999.875 Hz. Such peaks are present in most spectra and result from the beat of the reference light with unshifted illumination light scattered at e.g. the cell surface. The region between vertical lines is excluded (masked) in subsequent fitting and evaluation. b) Magnified central region of a particle-free signal measured in pure water (red) and 2.5% hydrogen peroxide solution (black). Inset: same data displayed for a larger frequency range. No signal, except for the sharp peak at the Bragg frequency (here at $\omega_B$ = 3000.3 Hz) emerging from a frequency independent noise floor, is detected. The spectral shape of these two successively taken spectra coincides near quantitatively

**Results and Discussion**

We now turn to various demonstrations of the data obtainable in SH-DLV experiments and the scope of the present instrument. First, we investigated the $SiO_2/TiO_2$ Janus particles (JP550) at a number concentration of $n_{JP550} = 5 \cdot 10^{14}$ m$^{-3}$ suspended in a 2.5% aqueous $H_2O_2$ solution. Their typical Doppler spectra with and without UV illumination are displayed in Fig. 3a). Without UV-illumination (red) a central Lorentzian indicates purely Brownian motion. With UV-illumination a slightly asymmetrically broadened Doppler peak, shifted to the right, indicates particle propulsion in the direction of **k**$_{ref}$ and allows for an easy discrimination between passive and active motions. This becomes even better visible in the inset of Fig. 3a showing a magnification of the central regions of the spectra.

At $\omega_B$ = 2999.875 Hz, a sharp line of full width at half maximum FWHM = $\Delta\omega_B \approx 0.1$ Hz is visible. Such a line is also present in the reference experiments performed in particle free water or 2.5% aqueous $H_2O_2$ solution (Fig. 3b) and most other spectra. We attribute it to parasitic scattering of illumination beam light from the static cell surface towards the detector, there beating with the reference beam exactly at the difference frequency $\omega_{B,\,exp}$. Its width and shape stay practically constant, while its position varies very slightly and slowly in time. For a typical half hour measurement we find $\omega_{B,exp}(t=0) = \omega_{B,nominal} \pm \Delta\omega = 3000.0\pm0.2$ Hz, and $\omega_{B,exp}(t) = \omega_{B,exp}(t=0)\pm0.04$ Hz. The temporal stability of the AOM therefore limits the spectral resolution to this value, which, however, is much smaller than the typical FWHM of the Lorentzian originating from particle diffusion, $\Delta\omega_{Diff} \approx 0.5$ Hz. In fact, the two successive runs shown in Fig. 3b coincide near quantitatively. In the subsequent evaluation of the spectra, we mask the data in the immediate neighbourhood of this peak, i.e. the eight points covering $\omega_B \pm 0.1$ Hz. The inset of Fig. 3a marks this central exclusion range by two vertical lines. Note further, that even in the UV-off case this still leaves enough points for fitting a Lorentzian with good accuracy (an example for such a fit is shown in Fig. 5b, below).

In turbid systems of decreasing transmission, the actual signal becomes drowned in a broad multiple scattering background. This sets the typical limit for signal detection to a transmittance of about 40% [41]. In Fig. 4a), a typical Doppler-spectrum of a moderately concentrated suspension (red data) and one taken in a multiple scattering environment (black data) is presented. Note the presence of a broad symmetric background in the latter signal, here obtained at a transmittance of 60%. To isolate the actual signal, we fit a Lorentzian (green solid line) to the outer wings of the broad background signal using the Levenberg-Marquardt algorithm implemented in Origin9.0 (OriginLab, Northampton, MA, USA), and subtract it from the data. Here we mask the immediate heterodyne signal region (2999.875 – 3002.0) Hz, and we perform the fit for the remaining frequencies in the interval 2990-3010 Hz.

Also the spectrum of the moderately concentrated sample is corrected, here by subtracting a constant background stemming, e.g., from detector shot-noise. In Fig. 4b), the background corrected SH-LDV spectra are shown for both cases. The signal shape due to the velocity distribution is now clearly recognized in both data sets. For the turbid case, the signal has become somewhat more symmetric. Both thus isolated single scattering signals can now be compared or subjected to further evaluation [42]. Here, we stick with a qualitative comparison. Both spectra in Fig. 4b show very similar shapes with the larger spectral power present for the sample with the larger concentration of (active) particles. The common centre of mass is at 3000.4 Hz with $\omega_B = 2999.875$ Hz. Note, that also in the turbid case an average positive Doppler shift is recorded. Here, the majority of passive particles contributes dominantly to the signal. The similarity of spectral shapes is, however, not compatible with active particles cruising through a stationary passive particle background. (For that case, one would expect a strong Lorentz-type spectral contribution centred about $\omega_B$.). This is therefore an interesting detail and suggests a JP motion induced local solvent motion carrying along the passive scatterers.

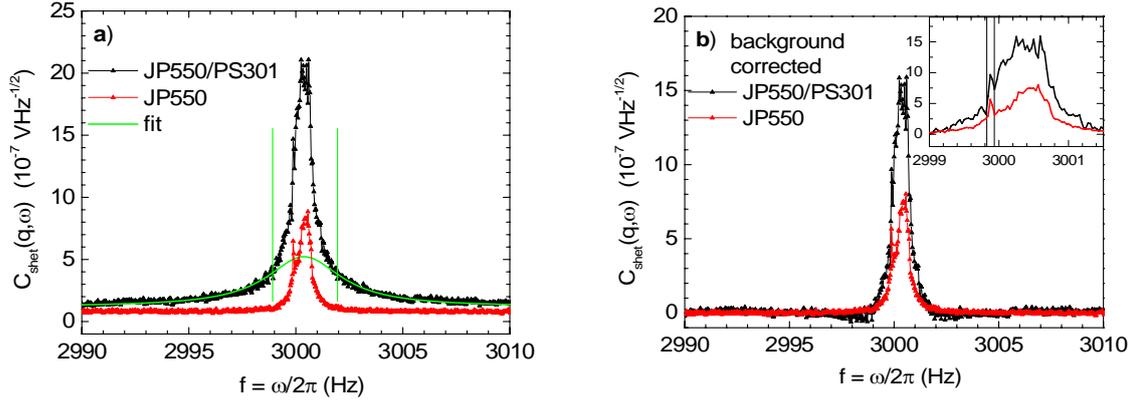

**Fig 4** Multiple scattering correction in SH-LDV. a) Comparison of raw data taken for JP550 at $n_{JP550} = 5 \times 10^{14}$ m$^{-3}$ in 2.5% hydrogen peroxide solution (red) and at $n_{JP550} = 2 \times 10^{15}$ m$^{-3}$ with the same peroxide concentration, but now in a suspension of passive PS310 at $n_{PS310} = 10^{16}$ m$^{-3}$ Note the presence of a broad background in the latter case, which increases symmetrically around the main signal. After masking the points between the vertical green lines, a Lorentzian (green) is fitted to the wings of this background signal and subtracted. This corrects for the multiple scattering contribution to the spectrum. In the first case, only a constant background is subtracted. b) Comparison of background corrected spectra. Now, the peak shape is clearly discriminated also for the case of the crowded, multiply scattering environment. Inset: Magnification of the immediate signal centre. Note that in both cases, the position of $\omega_B$ is clearly visible. While in subsequent fits of assumed velocity distributions, the region marked by the vertical lines should again be excluded, the location of $\omega_B$ can be used as reference corresponding to zero velocity. Here, both experiments show a positive Doppler shift.

Next, we demonstrate how to obtain quantitative information reliably from our SH-LDV instrument by comparing it to standard Dynamic Light Scattering (DLS). We recall that in SH-DLV we measure the Fourier transform of $g^{(1)}(q,\tau)$, retaining full information of both the individual particle velocity and diffusion. DLS, by contrast, measures the intensity autocorrelation function, $g^{(2)}(q,\tau)$. Using the Siegert-relation to obtain the coherent field autocorrelation func-tion, $g^{(1)}(q,\tau)$, the information on the absolute magnitude and on the direction of active or advective motion is removed (see, however, [36,34], who discuss the use of the incoherent, number density fluctuation related part of $g^{(1)}(q,\tau)$ for extracting information on velocity fluctuations in turbulent flow). In the coherent part, only changes in the mutual distances within pairs of scattering particle as caused by active, advective or diffusive motion are retained. The interpretation of the resulting decorrelation signal, however, requires a number of additional assumptions and even in simple cases, like linear simple shear, is extremely difficult (see e.g. [35]). Information on purely diffusive motion in a sample of homogeneous density and in the absence of advection can still be extracted from $g^{(1)}(q,\tau)$ obtained *via* Siegert's relation.

For measuring the diffusion constant of passive particles a long-time DLS measurement was done as benchmark for comparison to our method. Non-interacting JP550 (number concentration $n_{JP550} = 8 \times 10^{13}$ m$^{-3}$) suspended in pure water, *i.e.*, particles in the passive Brownian state in a singly scattering suspension were used. For precision determination of the diffusion coefficient by DLS, we follow the experimental protocol outlined in [48] assuring a sufficiently low noise level also for times long as compared to the principal relaxation time. The autocorrelation function extending over several orders of magnitude in time after the decay into the baseline was fitted with the 2$^{nd}$ order cumulant expansion under the assumption of purely Brownian motion (Fig. 5a).

Fig. 5b) shows the corresponding SH-LDV spectrum taken by our instrument at $n_{JP550} = 10^{14}$ m$^{-3}$. This spectrum was recorded in pure water with UV illumination on. Here we fitted a Lorentzian (red solid line) excluding the very central points about $\omega_B = 2999.875$ Hz again at a confidence level of 0.95. The inset of Fig. 5b shows a magnification of the central part of the spectrum; the vertical lines highlight the masked area. For the experiments shown in Fig. 5a) and 5b), we obtain $D_{SH-LDV} = (8.41 \pm 0.30) \times 10^{-13}$ m$^2$s$^{-1}$ and $D_{DLS} = (8.49 \pm 0.21) \times 10^{-13}$ m$^2$s$^{-1}$, respectively. Both values agree quantitatively within their uncertainties representing the standard error of the fits. From the averaged $D$, we obtain an average hydrodynamic radius of $a_h = (289 \pm 9)$ nm, somewhat larger than the dry SEM radius $a_{SEM} = (275 \pm 8)$. This type of discrepancy is a well-known effect that has been discussed extensively in the literature [8,49,50]. The excellent compliance with the theoretically expected curve shapes for these measurements comparing UV-off to UV-on conditions, moreover, shows that we cannot detect any convective out-of-plane motion. While this does not exclude the presence of UV-heating driven convection, it justifies its neglect in the interpretation of the spectra.

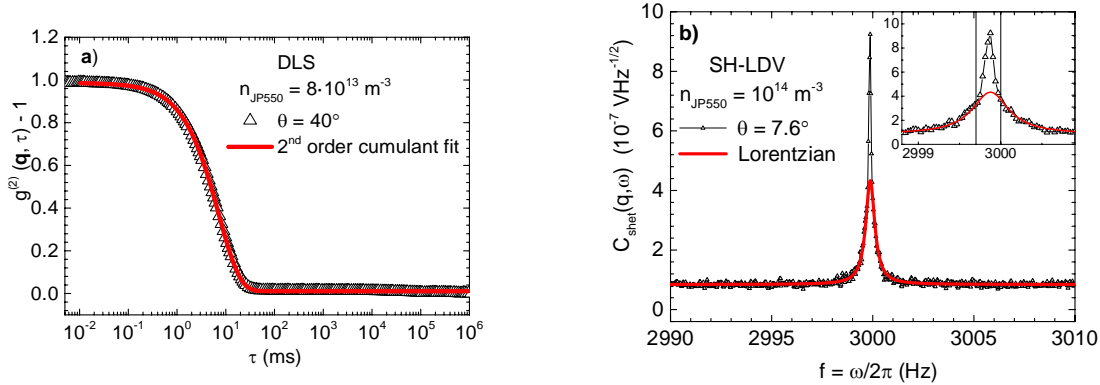

**Fig. 5** Determination of diffusion constants for particles suspended in pure water i.e. in the passive state. a) Autocorrelation function measured by Dynamic Light Scattering (DLS) of non-interacting JP550 at a number concentration of $n_{JP550} = 8 \times 10^{13}$ m$^{-3}$. A second order cumulant fit returns a diffusion constant of $D_{DLS} = (8.49 \pm 0.21) \times 10^{13}$ m$^2$s$^{-1}$. b) SH-LDV power spectrum of the same particles but with UV illumination on at $n_{JP550} = 10^{-14}$ m$^{-3}$. Fitting a Lorentzian to the wings of the experimental curve, and thus excluding the central static contribution (see inset) returns a diffusion coefficient of $D_{SH-LDV} = (8.41 \pm 0.30) \times 10^{-13}$ m$^2$s$^{-1}$.

The main purpose of SH-LDV, however, is the determination of velocities. For this, it is advantageous to switch between two different modes of data acquisition: frame by frame and continuous averaging. In sufficiently dilute systems, the former case measures small ensembles of individual particles, *i.e.*, their scattering contribution remains separated in frequency space. Examples of individual frames are shown in the stacked panel of Fig. 6a). Note the "noisy" appearance of the spectra due to summation over very few particles. This mode allows direct discrimination of runs that contain artificial perturbations when showing strongly deviating spectra, e.g., stemming from bubble formation or scattering by dust. Such data is readily distinguished and can be excluded from further evaluation. Moreover, any trend in velocities (e.g. a slowing from peroxide exhaustion or other systematic influences) will readily become observable this way. Hence, by employing the frame-wise acquisition mode we always assured that any systematic variation of velocities is caused by catalytic activity of the JP only. Fig. 6b) displays the corresponding histogram of observed velocities. The distribution is rather symmetric. Assuming a Gaussian distribution of velocities, the fit to the histogram returns an average velocity of v = 7.9 µm s$^{-1}$ and standard deviation normalized to the mean of σ = 0.23. We come back to this observation and its relation to the distribution of catalytic activities below.

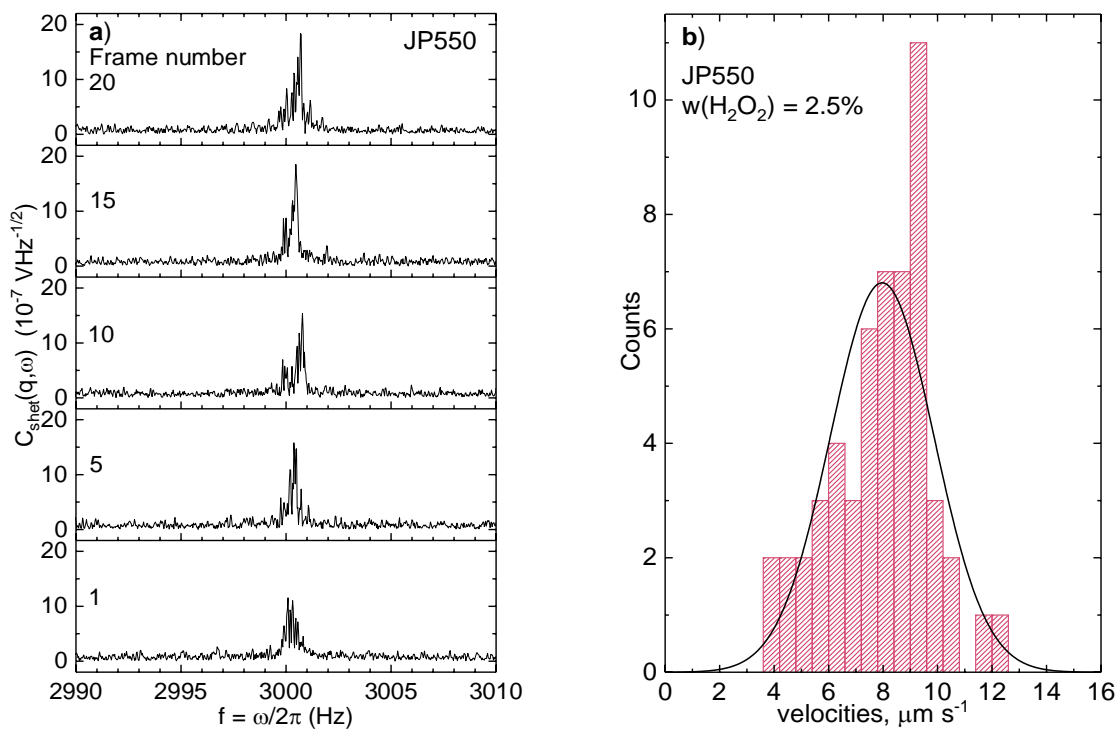

**Fig. 6** a) Selected individual frames collected for the actively propelling JP550 under UV illumination. Each frame duration is $t_{frame} = 32$ s with the overlap set to zero. b) Velocity distribution of 57 frame-wise averaged swimming speeds yielding a total average velocity of 7.9 µm s$^{-1}$. The solid curve is a fitted Gaussian assuming a normal distribution of catalytic activity with a standard deviation normalized to the mean of $\sigma = 0.23$. Given a homogeneous illumination and fuel distribution, and the absence of solvent flows, such a histogram corresponds to the distribution of average activities.

To prove the consistency between the aforementioned acquisition modes, the same data as in Fig. 6 is plotted again in Fig. 7a) and b) (down triangles). Fig. 7a) displays the typical smooth spectral shape obtained from continuous averaging, whereas in Fig. 7b) the velocities obtained from different acquisition modes are plotted frame-wise. The additional data in b) was taken during two subsequent runs, one averaging over $N = 57$ non overlapping frames (green circle) and the other ranging over $N = 47$ frames with 50% overlap (blue square). Using the latter parameters allowed for the reduction of the total measurement time. The average velocity of the frame by frame measurement coincides quantitatively with the velocities obtained in two independent runs using continuous averaging but different overlap conditions. Thus, if interested in obtaining only the average speed, the experiment duration can be considerably shortened. This also helps to avoid artefacts due to detection of unwanted long time effects like fuel exhaustion, coagulation and sedimentation.

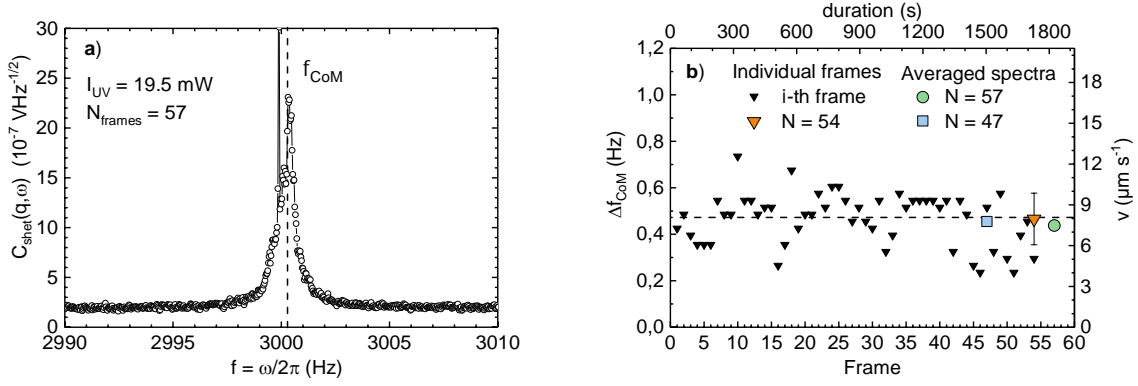

**Fig. 7** Quantitative determination of velocities at a peroxide concentration of $w(H_2O_2) = 2.5\%$. a) Power spectrum obtained by continuous averaging over 57 non-overlapping frames in one go. The difference between $\omega_B = 2999{,}9$ Hz and the centre of mass frequency $f_{CoM} = 3000.46$ Hz (dashed line) corresponds to a number averaged particle velocity $\bar{v} = 7.9\ \mu m\ s^{-1}$. b) Extracted velocities from different acquisition modes are shown to be within good agreement (Black down triangles: individual frame data; Orange down triangle: average of 54 individual frames; Light green circle: subsequent independent run over 57 successive frames without overlap. Light blue square: subsequent independent run over successive 47 frames with 50% overlap).

In Fig. 6b we had shown that the velocity distribution of frame-wise averaged spectra is only approximately Gaussian. We now discuss, how additional information may be extracted from this distribution and the spectral shape of the averaged spectra. In particular, we may identify the normalized distribution of average velocities (Fig. 6a) with the distribution of average activities, if certain assumptions are fulfilled: i) a homogeneous UV-illumination, ii) a homogeneous and constant peroxide concentration and iii) no significant additional particle flows inside the observation volume due to the locally induced collective swimmer flux. The first assumption is uncritical, since in our setup we only observe the homogeneously lit central region of the widened UV beam. The assumption of constant and homogeneous fuel concentration is difficult to test, but in principle could be feasible *via* additional fluorescence microscopy measurements using suitable dyes or by taking changes in pH as proxy. We performed preliminary measurements using micro-photometry [51] on JP550 particles in 3% peroxide solution in a horizontally placed microscopy cell lit homogeneously by UV light from below. Only a gradual change of overall pH but no significant spatial variation in pH was detected. This indicates a homogeneous distribution and consumption of peroxide, which however decreased in time.

The absence of convective solvent flows is the least reliable assumption. In fact, closer inspection of the spectra in Figs. 7a) and 8a) reveals that the distribution of velocities extends towards negative values. We attribute this finding to solvent convection induced by the locally enforced swimmer flux along the illumination direction. Due to solvent volume conservation, any local directed flow has to be accompanied by a corresponding backflow. Averaging over larger volumes will thus detect both flow types. Similar flows are well known from phoretic experiments in closed cells [41,42,44], but also from patterning experiments with light sensitive swimmers [52]. Due to the illumination/observation geometry in our experiment, solvent convection in negative z-direction along the cell walls will be monitored and readily yield a spectral contribution for frequencies smaller than $\omega_B$. Aligning the UV-laser counter-linear to the illumination beam $\mathbf{k}_{ref}$ and simultaneously increasing $\Theta$ should minimize this effect and further increase the magnitude of recorded Doppler shifts. Both expectations can be checked by analysing the corresponding spectral shapes. Thus, also the determination of activity distributions based on a large numbers of simultaneously measured UV-activated swimmers appears to be feasible with SH-LDV.

In the present experiment, UV-sensitive particles show motion away from the light source, i.e. motion along $\mathbf{k}_{UV}$ [14]. However, this can change under altered preparation conditions or be unknown for other types of self-propelling particles [6]. Therefore, it is a useful feature, to be able to discriminate explicitly the propulsion direction. SH-LDV offers this feature due to the fact that in the two branches of the spectrum the Doppler frequency enters as $\pm(\omega_B + \omega_D)$, *c.f.* Eq. (2). For the set-up shown in Fig. 2a), motion along $\mathbf{k}_{UV}$ translates to a positive Doppler shift with respect to $\omega_B$. By switching the angle of the incident UV light (with respect to the x-axis) from $+\alpha$ to $-\alpha$, we obtain a corresponding switch of Doppler shift, which is demonstrated in Fig. 8a. We obtain two spectra that are mirrored around the Bragg frequency $\omega_B$. Thus, upon change of the illumination direction, the projection of the swimming velocity onto the z-axis changes its sign. We here employed a cell with rectangular geometry well suited for directed motion in a unique majority direction. Other systems, like swarming bacteria, show different motion types like helical motion, run and tumble type motion or swarming. Here, it would be desirable to transfer the SH-DLV approach also to a goniometer-based instrument. First steps in this direction have recently been reported [44].

Another important point is the usefulness of SH-LDV for measurements under systematic changes of experimental boundary conditions and for their simultaneous control. Fig. 8b)

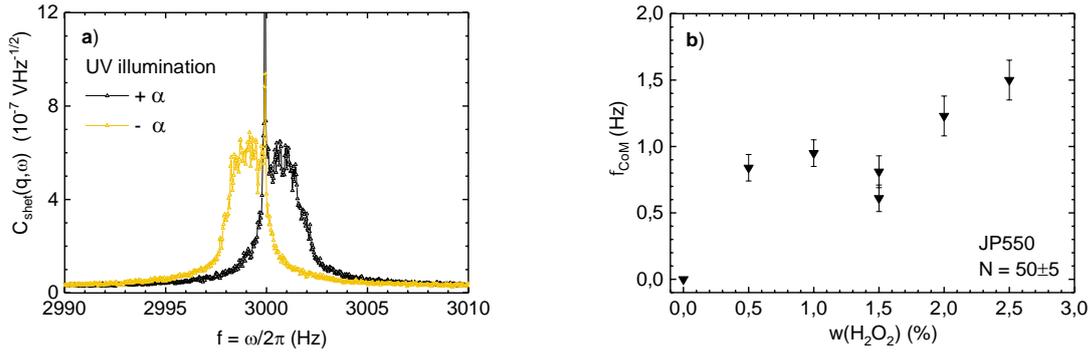

**Fig. 8** a). UV-activated JP550 Janus particles with mirrored incident direction of the UV laser beam as measured from the x-axis ($\alpha = +18.8°$ (black) and $\alpha = -18.8°$ (yellow)). Both spectra appear to be mirrored, with almost equal amplitudes, but Doppler shifts opposite in sign. The centre of mass frequencies are shifted by $\Delta f_{CoM}^{+15} = 0.65\ Hz$ and $\Delta f_{CoM}^{-15} = -0.72\ Hz$, respectively. b) Preliminary measurement on the $H_2O_2$ dependence of the swimmer velocity. The velocity increases with peroxide concentration. Note the large scatter of data and indicating the importance of precise control of experimental boundary conditions.

displays the results of (preliminary) measurements of the average swimming speed (here expressed in terms of the centre of mass Doppler frequency) as a function of peroxide concentration.

From literature, an increase of velocities with increased peroxide concentration is well documented [5,14], but its functional form for specific types of swimmers remains under debate. The data in Fig. 8b) was obtained by evaluation of continuously averaged spectra (N = 50 ± 5) for each peroxide concentration. In this set of experiments, the measurements were deliberately performed without any specific care about constant experimental boundary conditions. The UV-illumination intensity was approximately tripled and the peroxide concentrations were derived from added amounts of concentrated peroxide solution, but not checked via reaction with standard chemicals. Moreover, samples were investigated at different times after their preparation (sonicating a wafer piece and mixing the suspension with peroxide). As before, each run reliably determined the respective average swimming velocity, which increased as expected from literature. However, the data contains significant outliers, and a unique functional dependence cannot be inferred reliably. Both is attributed to systematic errors related to the reproducibility of experimental boundary conditions. Hence, due to its excellent statistics SH-LDV allows not only the precise determination of average swimmer velocities and directions but, moreover, permits immediate identification of insufficiently stabilized or less-controlled systematic investigations. Systematic measurements of the peroxide concentration dependence for differently sized Janus particles (including JP550) under well controlled boundary conditions are under way and will be reported elsewhere.

A particularly interesting boundary condition to be changed in a systematic way is the concentration of active particles. This was not yet tried here. In fact, we demonstrated the multiple scattering correction by adding passive particles in order to avoid the effects of particle interactions, which would involve several additional parameters into the interpretation. I. e. we kept the JP volume fractions below $10^{-4}$. Nevertheless, assuring otherwise well-defined boundary conditions, interaction effects should be discriminable and quantifiable in a systematic way. A general limit is set by the sample transmission for both UV illumination and laser light. UV absorption strongly depends on the choice of particle material. It can be made low, but a certain absorption to maintain catalytic activity has to be kept. Concerning the laser light, we have been able to measure average diffusion coefficients including standard error at a confidence level of 0.95 down to transmissions of $T \approx 0.4$. Without additional refractive index matching between particles and solvent, as well as on the particle size, this limit may be reached for $SiO_2/TiO_2$ JPs at volume fractions between 0.03 and 0.3.

For Janus swimmers, particle interactions come in three variants: hydrodynamic interactions, direct interactions and competition for fuel. All three depend on swimmer concentration. At large concentrations, one may expect the long ranged hydrodynamic interactions to lead to alignment effects as well as to phase-separation-like phenomena[28]. The first should be detectable for sufficiently large swarms via a reduction of diffusive broadening, the second by persistent large but random fluctuations of scattered intensity cross correlated in homodyne and heterodyne spectral contributions. Direct interactions of electrostatic origin are typically strongly screened in catalytic systems and are not expected to be easily discriminated. Moreover, their influence on the swimming dynamics is unclear. Since their main effect would be the introduction of an additional structural length scale, time resolved large angle scattering seems to be an alternative for their detection and characterization. Finally, competition for fuel may become an issue at either large densities or long times and lead to a significant slowing of swimmer motion. Here, it would be interesting to study the kinetics of slowing of average swimming speed for well defined initial conditions.

**Conclusion**

We presented SH-LDV as a facile and versatile approach to measure velocities of active systems in the bulk and away from a wall. In particular, it was shown to reliably measure self-propelling Janus particles while they undergo purely passive Brownian motion as well as in their catalytically active state. Passive motion was characterized by extracting the hydrodynamic radius. For the active swimmers we determined the direction, the average speed and the distribution of velocities. A determination of the activity could be obtained additionally. Most importantly, SH-LDV can characterize the swimming performance not only on isolated, non-interacting swimmers, but also in bulk environments crowded by passive particles. Working in frequency space allows us to remove the multiple scattering contributions occurring in turbid systems by a facile correction scheme. This leads to a statistically trustworthy method, highly sensitive to environmental changes. The system chosen for demonstration were photocatalytic active Janus particles of $SiO_2/TiO_2$ type. However, the method is not restricted to this specific system, and we anticipate that SH-LDV opens access to characterize the swimming properties for a wide variety of swimmer classes and types.

**Acknowledgements**

Financial Support of the DFG in the priority program SPP 1726 (Grant Nos. Pa 459/18-1,2 and FI 1966/1-1,2) is gratefully acknowledged. We thank Prof. T. Sottmann for valuable discussions.

**Conflict of Interests**

The authors declare that they have no conflict of interest.